# Exploring Novelty Differences between Industry and Academia: A Knowledge Entity-centric Perspective

Hongye Zhao [a], Yi Zhao [b], Chengzhi Zhang [a,*]

[a] Department of Information Management, Nanjing University of Science and Technology, Nanjing, 210094 China
[b] School of Management, Anhui University, Hefei, 230601, China.

**Abstract:** Academia and industry each possess distinct advantages in advancing technological progress. Academia's core mission is to promote open dissemination of research results and drive disciplinary progress. The industry values knowledge appropriability and core competitiveness, yet actively engages in open practices like academic conferences and platform sharing, creating a knowledge strategy paradox. Highly novel and publicly accessible knowledge serves as the driving force behind technological advancement. However, it remains unclear whether industry or academia can produce more novel research outcomes. Some studies argue that academia tends to generate more novel ideas, while others suggest that industry researchers are more likely to drive breakthroughs. Previous studies have been limited by data sources and inconsistent measures of novelty. To address these gaps, this study conducts an analysis using four types of fine-grained knowledge entities (Method, Tool, Dataset, Metric), calculates semantic distances between entities within a unified semantic space to quantify novelty, and achieves comparability of novelty across different types of literature. Then, a regression model is constructed to analyze the differences in publication novelty between industry and academia. The results indicate that academia demonstrates higher novelty outputs, which is particularly evident in patents. At the entity level, both academia and industry emphasize method-driven advancements in papers, while industry holds a unique advantage in datasets. Additionally, academia-industry collaboration has a limited effect on enhancing the novelty of research papers, but it helps to enhance the novelty of patents. This framework overcomes the limitations of literature type, providing a generalizable tool for comparing novelty between academia and industry. We release our data and associated codes at https://github.com/tinierZhao/entity_novelty.

## Introduction

Academic research focuses on theoretical inquiry and the advancement of fundamental science, aiming to expand human knowledge and drive disciplinary progress (Merton, 1973; Sauermann & Stephan, 2010). In contrast, the industrial sector emphasizes core competitiveness (Geisler, 1995), prioritizing economic returns and often safeguarding intellectual appropriability by restricting the disclosure of research outcomes (Chirico et al., 2020).

Following this logic, the industry would typically choose to limit the disclosure of novel research outcomes to safeguard its competitive advantage. However, this traditional notion is being challenged in the field of artificial intelligence (AI). Although different from the institutional positioning of open knowledge sharing in academia, the industry actively explores openness while maintaining core competitiveness (Zhao et al., 2026). For example, leading tech companies have released cutting-edge technologies in algorithms and models, such as the BERT model (Devlin et al., 2019) and various other open-source large language models. Additionally, they have significantly lowered the barriers to adopting artificial intelligence technologies by offering application programming interfaces (APIs) and detailed technical documentation. Moreover, the industrial sector's active participation in the most active and popular AI conferences has spurred numerous disruptive innovations (Liang et al., 2024). While these measures reduce knowledge appropriability, they yield significant benefits: attracting developer communities to cut maintenance costs while generating revenue through technical services, and enabling access to external knowledge via collaborations to sustain technological leadership, foster product optimization through knowledge spillovers (Jiang et al., 2024), and expand market share.

This raises an unresolved question: do industry research outcomes exhibit lower novelty than academia's? Currently, scholars are still divided on the issue. Some scholars argue that academia contributes more novel ideas, while industry tends to adopt and refine academic advancements (Bikard & Marx, 2020). Subsequent studies further confirm academia's leadership in AI innovation (Liang et al., 2024). However, Dwivedi et al.

---

(2021) suggest that industry researchers are more likely to drive new AI technologies. The rise of pre-trained models such as Transformer and GPT, along with the rapid development of large-scale language models like ChatGPT, Ahmed et al. (2023) highlights the industry's dominance in computational resources, data, and talent. As a key branch of artificial intelligence, NLP has witnessed a surge of groundbreaking achievements from both academia and industry, making it a typical scenario for exploring this question.

Patents and papers are the main scientific and technical outputs in academia and industry. Patents carry technology and business knowledge, and their text analysis is an important tool for technology development and management (Arts et al., 2021; Lee et al., 2009). As the core of basic research, papers are the key medium for academic exchanges (Ba et al., 2024). This study measures the novelty of industry and academia through patents and papers. However, evaluating the novelty of scientific and technical literature is inherently challenging. Multiple studies have confirmed that highly innovative research is difficult to identify during peer review (Koppman & Leahey, 2019; Liang et al., 2023; Riera & Rodríguez, 2022; Wang et al., 2017). In addition, the current research on novelty in AI primarily focuses on papers (Chen et al., 2024; Liang et al., 2024) and fails to comprehensively consider various scientific and technical literature such as patents. This research limitation does not stem from issues of data availability but rather from differences in the methods used to evaluate novelty between patents and scientific papers. For scientific papers, novelty is typically measured through journal citation pair analysis (Lee et al., 2015; Uzzi et al., 2013; Wang et al., 2017). Moreover, the classification codes commonly used in patents cannot be aligned with those used in scientific papers (Verhoeven et al., 2016), making it impossible to reconcile the two. Additionally, some studies have attempted to conduct evaluations based on sentence-level semantics (Jeon et al., 2022, 2023), but sentence semantic vectors are easily influenced by academic writing styles, and the writing styles of patents and papers differ significantly, resulting in incompatibility with this method.

By further focusing on knowledge elements in the literature, precise alignment of measurements can be achieved through refining the granularity of their representation and relying on a unified conceptual space constructed by word embedding models. As noted by Aceves and Evans (2024), concepts and conceptual spaces serve as crucial knowledge foundations for the development of numerous organizational theories. Even when knowledge is expressed in different forms across various scenarios, the semantic connections and differences among them can be quantified through a unified word embedding model (Aceves & Evans, 2024).

This study addresses the gap by using a unified novelty evaluation framework that leverages fine-grained knowledge entities to assess the novelty of publications across academia, industry, and their collaborations in NLP. Specifically, we map these fine-grained knowledge entities to a unified conceptual space (Aceves & Evans, 2024), calculate the semantic distances between fine-grained knowledge entities, and assess the difficulty of different entity combinations.

Specifically, we address the following three research questions:

*RQ1*: How to unify the novelty calculation method based on fine-grained knowledge entities for both papers and patents?

*RQ2*: How do novelty manifestations differ across entities in various institutions and document types?

*RQ3*: Is there a difference in the novelty of scientific and technical literature between industry and academia?

The contributions of this paper are as follows:

First, this study extracts knowledge entities from patents and papers, mapping both to a unified semantic space constructed by SciBERT. By calculating the semantic distance between entity pairs to quantify combination difficulty, this framework preserves fine-grained interpretability at the entity level while providing a foundation for novelty comparison across documents.

Second, existing research is constrained by relying on a single type of literature, and conclusions remain inconsistent. This study integrates top conference papers with the United States Patent and Trademark Office (USPTO) patent dataset to achieve reliable validation of the research hypotheses.

The code and data used in this study are open-sourced on GitHub and can be accessed via the following website: https://github.com/tinierZhao/entity_novelty.

**Related work**

For the research questions proposed in this paper, we conducted a review of the scientific and technical literature on novelty measures, as well as the factors influencing novelty.

*Novelty measures in the scientific and technical literature*

The measurement of novelty helps identify valuable innovations in advance and provides key insights for technological transfer and innovation. Currently, novelty is primarily measured through combinations, as

Nelson and Winter (1985) argued, "the creation of novelty mainly involves the recombination of existing conceptual and physical materials". Traditional methods for measuring novelty include the use of journal pairs and classification code pairs to assess the novelty of literature. With the availability of large-scale data and the advancement of machine learning and natural language processing technologies, novelty measurement methods have been continuously innovated. The combination of other types of knowledge elements has gradually become an important approach for assessing novelty. Additionally, some studies have explored new avenues by treating novelty as a binary classification task, using classification or outlier detection methods to distinguish between novel and non-novel literature.

From a combination-based view, early methods primarily focused on citation references and classification codes. Uzzi et al. (2013) compared the observed and Monte Carlo-simulated frequencies of journal pairs to calculate the z-score for each pair, using the lowest 10th percentile z-score to indicate a paper's novelty and the median z-score to indicate its conventionality. Lee et al. (2015) improved Uzzi's method in terms of computational difficulty by adopting a multi-year time window, which reduced the previous single-year window and calculated the commonness of citation pairs. Wang et al. (2017) measured novelty through the first-time combination of different citation journal pairs in a paper. Specifically, they constructed a co-citation matrix for the journals and used cosine similarity between the vectors of each journal to assess the difficulty of combining the journal pairs.

Regarding patent novelty measurement, early traditional methods focused on patent classification codes and backward citations (Lee & Lee, 2019; Verhoeven et al., 2016). However, citations merely describe existing technologies and fail to reflect the technology of the patent itself, often presenting incomplete and biased representations (Arts et al., 2021). Measuring technological novelty through patent IPC codes (Fleming, 2001) is overly broad and tends to capture interdisciplinarity rather than technological uncertainty.

With the continuous development of NLP technologies, tasks such as scientific terminology extraction (entities, keywords) and semantic embedding have matured, measuring novelty based on scientific text content is a more reasonable and effective approach. Liu et al. (2022) used the BioBERT model to calculate the semantics of biological entities, determining entity pair novelty based on semantic similarity. The novelty score for each paper is calculated as the proportion of novel entity pairs to the total possible entity pairs. Similarly, Chen et al. (2024) applied an entity similarity-based approach to evaluate the novelty of conference papers in NLP. Luo et al. (2022) employed BERT word embeddings to measure novelty by assessing the novelty of research questions, methods, and their combinations. Arts et al. (2021) extracted keywords from patent titles and abstracts, calculating "new_ngram" and corresponding "new_ngram_reuse" to measure patent novelty. Wei et al. (2024) used the BERT model to extract innovative sentences from patent claims and distilled them into knowledge element triples, measuring novelty scores for the triples by projecting entities and relations into a common space. Shi and Evans (2023) measure research novelty by leveraging hypergraph embeddings to capture high-dimensional content correlations. Their findings reveal that breakthroughs more frequently emerge from "knowledge expeditions" rather than traditional interdisciplinary teams.

From the perspective of binary classification, Jang et al. (2023) treated patent novelty as a classification task, using RoBERTa for semantic embedding of patent claims to develop a self-explainable novelty classification model. Jeon et al. (2022) embedded patent claims and used the local outlier factor (LOF) algorithm to calculate patent novelty. Their study showed that, although ELMo and BERT provide high-quality patent embedding vectors, they are less suitable for modeling the technological features of patents, particularly in single technical domains, compared to Doc2Vec. X. Liu et al. (2025) constructed an unsupervised learning framework using the Doc series model and the LOF. Jeon et al. (2023) trained a FastText model using paper titles in the biomedical field and applied the LOF algorithm to measure the novelty score of each paper.

From the above-mentioned studies, the methods for measuring novelty have evolved from the early approaches relying on citation and classification codes to those based on text content analysis. However, there are still shortcomings in evaluating novelty. First, early methods based on citations (Lee et al., 2015; Uzzi et al., 2013; Wang et al., 2017) and patent classification codes (Fleming, 2001; Lee & Lee, 2019; Verhoeven et al., 2016) suffer from inherent flaws: citation metrics are affected by self-citation bias, incomplete records, and may confuse "interdisciplinarity" with "novelty" (Fontana et al., 2020), and the computational cost of analyzing large-scale data is proliferating and inefficient. Patent classification codes are overly granular and difficult to align with papers. Although text-based methods have improved the granularity and interpretability of measurements in recent years, existing research mostly focuses on a single field (or papers or patents) and lacks a framework for uniformly assessing the novelty of the two.

In the following chapters, fine-grained knowledge entities will be uniformly extracted from papers and patents, mapped to a shared conceptual space constructed based on word embeddings, and novelty will be evaluated by calculating semantic distance. This framework not only retains the interpretability and fine-grained advantages of the substantive method but also provides a unified basis for cross-domain novelty comparison between papers and patents, filling the gap in current evaluation tools.

*Factors influencing the novelty of scientific and technical literature*

Previous studies have explored the relationship between novelty from various perspectives, including institutional nature, team size, and author attributes within teams.

Regarding team size, existing research presents inconsistent findings. Uzzi et al. (2013) found that research teams are more likely to introduce novel combinations within familiar knowledge domains compared to single-author papers. Lee et al. (2015) identified an inverted U-shaped relationship between team size and novelty, with this effect largely driven by the interplay between team size and knowledge diversity. Wu et al. (2019) suggested that smaller teams are more likely to disrupt science and technology with new ideas, while larger teams tend to focus on existing ones. Shin et al. (2022), using Web of Science data, found that scientific collaboration negatively affects novelty, as collaborative research tends to remain within established fields. However, Wu et al. (2024) argued that collaboration fosters trust and problem-solving abilities, and that knowledge diversity enhances knowledge transfer and promotes the impact of science on technology. Conversely, some studies indicate that excessive team heterogeneity may reduce trust, hinder knowledge sharing, and obstruct innovation (Chen et al., 2015).

At the author attribute level within teams, teams with diversified expertise tend to produce more original work and have a long-term advantage in terms of impact (Zheng et al., 2022). Mori and Sakaguchi (2020) examined how differentiated knowledge among inventors enhances patent novelty using Japanese patents. Gender diversity within teams has also become a favored topic in recent years. Teams with gender diversity produce papers with higher novelty and greater impact compared to single-gender teams (Yang et al., 2022). Liu et al. (2024) explored the relationship between novelty and gender heterogeneity in doctoral theses, finding that female authors had lower average novelty scores than male authors, and male advisors were more likely to supervise students who produced theses with higher novelty. Notably, this gender difference was more pronounced in lower-prestige universities. Similarly, Chan and Torgler (2020) found that among elite scientists, female scientists tend to receive more citations than their male counterparts.

At the institutional level, academia tends to lead industry in terms of novelty at the paper level, generating more exploratory ideas, while industry is more likely to produce high impact papers (Liang et al., 2024). Chen et al. (2024) measured the novelty in the NLP field, finding that academia and collaborative institutions tend to be more novel than industry, based on fine-grained combinations of knowledge entities. Other studies suggest that papers involving companies have a higher impact, and collaborations between industry and academia exhibit greater novelty (Jee & Sohn, 2023).

Chen et al. (2024) share similarities with this study in exploring novelty within the NLP domain. However, this research further incorporates patent data to compare novelty performance between academia and industry across both types of literature. It is important to note that this is not merely a data expansion exercise. Instead, it is grounded in a unified bottom-up conceptual space, leveraging word embedding models to establish comparability in semantic distances between entities across different types of scientific literature (Aceves & Evans, 2024). This approach offers greater versatility, transcending limitations imposed by document types and writing styles, and demonstrates superior scalability. Compared to Chen et al. (2024), which focused solely on entity frequency distribution and type proportions, this study further analyzes trends in semantic distance variations across different institutions within various innovation categories. Additionally, we comprehensively address the issue of "diminishing marginal returns in innovation with increasing component count" (Fleming & Sorenson, 2001), which was not explored in Chen's research. Furthermore, our findings diverge from theirs: we observe that academia-industry collaboration has a limited impact on enhancing paper novelty, consistent with the conclusions of Liang et al. (2024).

This study examines the novelty performance of different institutional types in patents and papers. Currently, research on institutional types at the substantive level remains scarce and is largely confined to single datasets. Although academia and industry produce abundant papers and patents, respectively, few studies have employed a comparative framework to assess the novelty of these two sectors across multiple types of scientific literature.

**Methodology**

This study aims to quantify the performance of different team compositions on the novelty of scientific and technical literature. The research framework in Figure 1 outlines four key phases:

Phase 1. Dataset construction: We constructed an original dataset that includes scientific and technical literature in the NLP field, comprising papers and patents published between 2000 and 2022.

Phase 2. Entity extraction: Fine-grained knowledge entities were extracted from both scientific papers and patents, with the knowledge from scientific papers being transferred to patents.

Phase 3. Novelty measurement: A unified pre-trained model is applied to obtain the semantic vectors of each extracted knowledge entity. The difficulty of combining these entities is measured based on the semantic distance between them (Chen et al., 2024; Liu et al., 2022, 2024). We then use this semantic distance-based method to evaluate the novelty of each document.

Phase 4. Regression modeling: Conducted a combined regression analysis on the standardized data to reveal the overall novelty trends across different literature types. Subsequently, separate regression analyses were performed for patents and papers as a heterogeneity test to examine whether the observed trends varied by literature type.

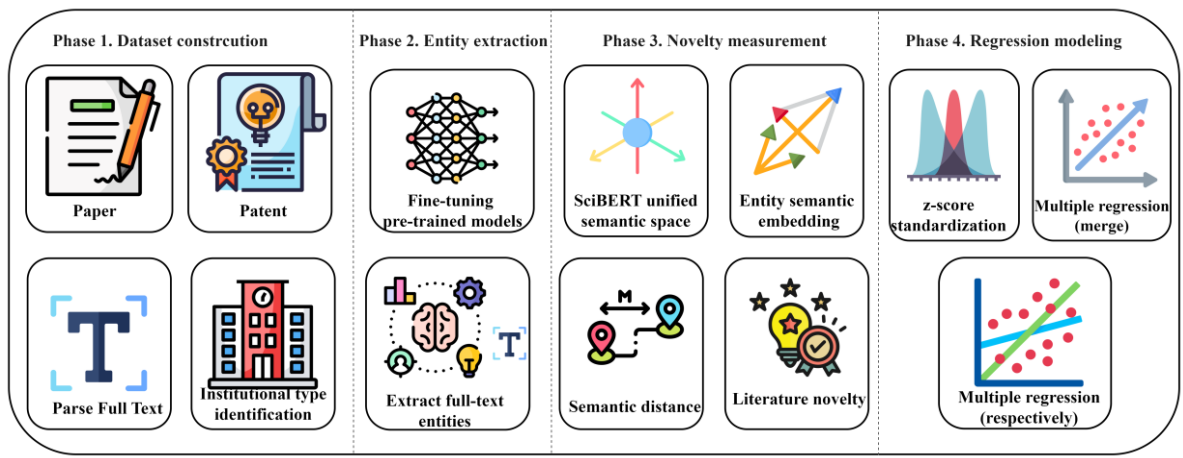

**Figure 1. Framework of this study**

*Data collection*

Both paper and patent datasets of this study are derived from the unified NLP field. Paper data was collected from the ACL Anthology[1] website. We selected three representative conferences for our study: ACL (Annual Meeting of the Association for Computational Linguistics), EMNLP (Conference on Empirical Methods in Natural Language Processing), and NAACL (North American Chapter of the Association for Computational Linguistics). A total of 17,783 full-text papers from 2000 to 2022 were collected.

The patent data was collected from the United States Patent and Trademark Office (USPTO) through the PatSnap[2] system. We conducted a search for patents within the time frame of 2000 to 2022, using the following query: CPC:(G06F40*[3]) AND APD:[20000101 TO 20221231] AND COUNTRY:US. We focused on invention patents and filtered out those with legal statuses such as withdrawal, rejection, abandonment, application termination, or complete invalidation. Additionally, patents with the same priority were consolidated into families. Ultimately, a total of 25,305 patents were obtained.

*Institutional type classification*

For the classification of authors' institutional types in papers, we built on the work of Chen et al. (2024), who manually integrated the Global Research Identification Database (GRID)[4]. We further supplemented the missing institutional data for papers in our database by combining manual annotation with OpenAlex queries, aiming to address the issue of missing institutional information for papers (Zhang, Cao, et al., 2024). It should be noted that OpenAlex adopts the same institutional classification standards as GRID, with both being based on the Research Organization Registry [5] (ROR) system. The system includes the following types of institutions: education, company, facility, nonprofit, government, healthcare, archive, and other. Finally, we completed the author information and their corresponding institutional affiliations for 17,783 papers. For each author, following previous research, we considered the first-listed institution as the author's primary affiliation (Hottenrott et al., 2021).

---

[1] https://aclanthology.org/

[2] https://www.patsnap.com/

[3] CPC: G06F40, Handling natural language data

[4] The Global Research Identification Database (GRID) classifies institution types into eight categories: government, education, company, facility, healthcare, nonprofit, archive and other. GRID collects institution names globally and supplements institution data, such as country and city of location, using Wikipedia (www.wikipedia.org) and GeoNames (www.geonames.org).

At the institutional classification level, given that this study aims to explore novelty differences between the academic and industrial sectors, we focused on two categories: academic institutions and industrial institutions. Specifically, institutions categorized as "company" were classified into the industrial sector, while those categorized as "education" were classified into the academic sector. Additionally, in line with the definition criteria from existing studies (Chen et al., 2024; Xu et al., 2022), for healthcare-type institutions, the main purpose of their publications is research-oriented, so we classify educational and healthcare institutions into the academic category. As for other types of institutions, they are not covered in this study, so we ignore them.

Specially, a paper is classified as "Academia" if all its authors are affiliated with academic institutions (such as universities or research institutes), as "Industry" if all authors are affiliated with industry institutions (such as companies or corporations), and as "Cooperation" if it involves authors from both academia and industry. If the paper contains neither industry authors nor academic authors, it is categorized as "Other".

The patent data processing begins with extracting standardized applicant information from databases, where all non-personal names are presented in either Chinese or English. An Edit distance algorithm (Levenshtein, 1965), combined with a local dictionary, is then applied to normalize institutional names. Based on lexical features, two sets of keywords were defined: one for academic institutions and one for industrial organizations, covering both English and Chinese terms. The algorithm classifies institutions containing education-related terms (such as "edu", "univer") as academic, and those with company-related terms (such as "inc", "ltd", "lp") as industrial. This method ensures efficiency and accuracy, as the database provides standardized applicant fields. For unrecognized institutions, spaCy[6] named entity recognition is used to determine whether the applicant is "Individual". For individual applicants appearing more than twice, we validate with ChatGPT to check for missed categorizations from industry or academia (prompt see Appendix Table 5). Those confirmed as individual applicants through ChatGPT verification will be categorized as "Individual". All unidentified institutions that cannot be assigned to academic, industrial, or individual categories through the above process are defined as "Other". Finally, the results are manually reviewed to correct and supplement the algorithm's output.

The specific institutional distribution for papers and patents is shown in Table 1.

Table 1. The institutional distribution of scientific and technical literature

| Institution Types | Count | Ratio (%) | Count | Ratio (%) |
|---|---|---|---|---|
| | Paper | | Patent | |
| Academia | 11,965 | 67.28 | 470 | 1.86 |
| Industry | 1,409 | 7.92 | 21669 | 85.63 |
| Cooperation | 3,652 | 20.54 | 71 | 0.28 |
| Individual | 0 | 0 | 2932 | 11.59 |
| Other | 757 | 4.26 | 163 | 0.64 |

*Knowledge entity extraction and normalization*

This study draws on the classical theory of combinatorial novelty (Fleming, 2001; Uzzi et al., 2013), measuring novelty based on atypical combinations of knowledge components. This degree of atypical could be quantified by calculating semantic distance, a method widely adopted in bibliometrics (Chen et al., 2024; Liu et al., 2022, 2024). Considering that most natural language processing research typically encompasses the following key elements: 1) dataset construction or selection, often involving text resources such as corpora and dictionaries, which serve as the foundation for model training and validation; 2) method selection and application, which defines the strategies and steps for solving problems; 3) the choice of evaluation metrics, used to measure model performance and task quality; 4) the use of tools, including programming languages, software, and open-source tools required for implementing and testing NLP methods (Pramanick et al., 2025; Zhang, Zhang, et al., 2024). Based on this framework, we extract fine-grained knowledge entities from each patent and paper, covering the categories of Method, Tool, Metric, and Dataset.

---

[5] https://ror.org/
[6] https://pypi.org/project/spacy/

In the fine-grained knowledge entity recognition task, we used the pre-trained SciBERT model. Due to differences in writing style and text structure between patents and papers, we trained separate entity recognition models for each type of document. Specifically, for papers, we adopted the framework proposed by Zhang et al. (2024). For patents, we initially applied a pre-trained model to annotate the patent texts, followed by re-annotation of the extracted entities according to the labelling rules. Additionally, for unique entities in patent texts, such as Storage medium, we performed extra annotation. After several rounds of iteration and adjustments, we obtained the patent entity recognition model (SciBERT + CRF), which achieved the following performance: Precision of 78.83%, Recall of 82.51%, and F1 score of 80.63%. Given that extracting entities only from titles and abstracts would miss many, we performed full-text extraction for both patents and papers. Paper data were extracted from PDFs, and the patent database was also exported in full text. For entity normalization, we used Edit distance (Levenshtein, 1965) to cluster entities. Ultimately, we identified 37,236 entities in the papers and 9,523 entities in the patents.

Appendix Table 4 presents the top 5 entities in each category for both patents and papers. Since patents are rarely evaluated on public datasets, the proportion of Dataset entities in patents is quite low, and as a result, the recognition performance for these entities is somewhat weaker. Additionally, a distinctive feature of patent terminology is its level of abstraction, particularly evident in the claims section. Unlike general discourse, which relies on precise wording to accurately convey content and avoid vague or overly broad terms, patent claims intentionally use generalized vocabulary (Codina-Filbà et al., 2017). This strategy enables companies to broaden the scope of their intellectual property protection, ensuring more extensive exclusivity over their innovations (Ashtor, 2022). Furthermore, descriptions of Tool entities in patents tend to be more generalized, reflecting this situation.

*Semantic distance-based novelty measurement*

We explore the novelty of entity combinations through an analysis of the fine-grained knowledge entities extracted from scientific and technical literature. We draw on the work of Liu et al. (2022, 2024) in the field of scientific novelty assessment for biomedical papers. They treated biological entities as core elements of the research method and used the pre-trained Bio-BERT model to quantify the semantic distance between these entities to measure novelty. We applied this approach to evaluate the novelty of papers and patents in NLP, using pre-trained SciBERT to calculate the semantic similarity of entities for novelty measurement.

Specifically, when using SciBERT for embedding processing, we analyze its built-in [CLS] token. This token serves as the starting point of a sentence and aggregates the contextual information of the entire sentence, thus serving as the vector representation of the entire sentence. Based on this characteristic, we treat patent entity phrases as short sentences and obtain the semantic vector representation of the patent entity by outputting the vector at the [CLS] position. To concretize the semantic vector representation, we take the NLP core entity "transformer" as an illustration: we input the entity to SciBERT in the standardized format, and the model outputs a 768-dimensional dense vector via the [CLS] token. The first 5 dimensions of this vector are presented below: [3.70797575e-01, 1.42690563e+00, -8.87082458e-01, -5.18282456e-03, 5.78712046e-01, … (subsequent 763 dimensions omitted)]

For an entity pair $(e_i, e_j)$, the distance between the two is denoted as $D$, and $cosine(e_i, e_j)$ represents the semantic similarity between the entities.

$$D(e_i, e_j) = 1 - cosine(e_i, e_j) \tag{1}$$

We marked the top 10% of entities with the high semantic distance as high novelty entities. Finally, we analysed the frequency of these high novelty entities in the text and measured the novelty of each paper based on their proportion in all entity combinations.

*Analysis of novelty differences*

To investigate the differences in novelty across various institutions, this study employs regression analysis to quantify and compare the novelty demonstrated in the scientific and technical literature produced by different institutions. The following sections provide a detailed description of the process of variable selection and the construction of the regression model.

Dependent variables: In the setting of independent variables, we first use the continuous novelty indicator (Novelty Score) calculated in the previous section for analysis. This score measures the proportion of novel entities in each paper or patent, valued between 0 and 1, where a higher score signifies greater novelty. It enables comparison of mean novelty output differences between industry and academia. Second, to address potential uncertainty in novelty outcomes and validate conclusion robustness, we define the top 10% of papers/patents ranked annually by this score as "high novelty" and construct a binary variable (NS Top) (Jeon et al., 2022), coded 1 for the top 10% and 0 otherwise. This variable primarily serves to analyze which type of sector is more likely to generate high novelty papers/patents.

Independent variables: This study defines the independent variables as the type of institution. After excluding institutions categorized as "other" and "individual", the remaining institutions are classified into three categories: academia, cooperation, and industry. Specifically, two binary variables (Academia and Cooperation), are defined. The Academia variable is set to 1 if the literature belongs to an academic institution, and the Cooperation variable is set to 1 for literature from cooperative institutions, with both variables set to 0 for literature from industry.

Control variables: To eliminate the interference of other factors and identify the net effect of institutional type on novelty, this study included multiple control variables. First, team-related variables were considered, including the number of institutions (Institutions num) and the number of authors or inventors (Au/In num). This is because large R&D teams typically possess broader and more extensive knowledge bases, and collaboration among team members can increase opportunities for cross-disciplinary knowledge integration (Wu et al., 2025). Additionally, such teams tend to conduct developmental research (Wu et al., 2019), and their characteristics may influence novelty. For patents, we also include the size of the patent family (Family size), which is commonly associated with welfare value and technological impact (Kabore & Park, 2019). Furthermore, the number of IPC classification codes at the subgroup level (IPC num) is controlled to account for the diversity of the patent's knowledge components (Sun et al., 2022). Additionally, considering that this study calculates novelty through entity combinations, the number of entities (Entity num) is also included as a control variable to eliminate the interference of entity quantity. Furthermore, year dummy variables were introduced to control for potential annual differences that could influence the results. For continuous control variables, following the method of Wu et al. (2024, 2025), a log transformation was applied to address their skewed distributions.

When performing regression, we only retained scientific literature from industry, academia, and academia-industry collaborations. In addition, we excluded outliers: entries with fewer than five entities and novelty scores of zero. The final dataset contained 16,295 papers and 20,934 patents. The summary statistics of the variables and the correlation coefficients between the variables are presented in Table 2 and Figure 2, respectively.

We found a strong correlation between the continuous and discrete forms of the dependent variable (novelty), while the correlations between the independent and dependent variables were weak. We then calculated the variance inflation factors (VIFs) for all explanatory variables to assess multicollinearity. The VIF for papers was 2.30 and for patents was 1.06, both below the threshold of 5 (Marcoulides & Raykov, 2019). These results indicate that multicollinearity has minimal impact on our model, ensuring the reliability of the estimates.

**Table 2. Summary statistics of variables for regression analysis (N = 20,934 patents, N = 16,295 papers)**

| *Variable* | **Paper** | | | | **Patent** | | | |
|---|---|---|---|---|---|---|---|---|
| | *Mean* | *Std. Dev.* | *Min* | *Max* | *Mean* | *Std. Dev.* | *Min* | *Max* |
| Novelty Score | 0.10 | 0.06 | 0.01 | 0.52 | 0.12 | 0.07 | 0.01 | 0.67 |
| NS Top | 0.10 | 0.30 | 0 | 1 | 0.10 | 0.30 | 0 | 1 |
| IPC num | - | - | - | - | 1.95 | 1.00 | 1 | 10 |
| Family size | - | - | - | - | 2.08 | 1.94 | 1 | 82 |
| Au/In num | 3.82 | 2.24 | 1 | 77 | 3.30 | 2.13 | 1 | 26 |
| Institutions num | 1.84 | 1.24 | 1 | 44 | 1.05 | 0.42 | 1 | 15 |
| Entity num | 34.93 | 14.53 | 8 | 153 | 29.81 | 14.95 | 8 | 98 |
| Academia | 0.70 | 0.46 | 0 | 1 | 0.02 | 0.14 | 0 | 1 |
| Cooperation | 0.22 | 0.41 | 0 | 1 | 0.00 | 0.05 | 0 | 1 |

**Note:** The papers do not include IPC numbers or Family size, which are represented as "-".

Then multivariable regression was conducted to examine how different types of institutions influence the novelty scores of the literature.

$$Novel_i = \alpha + \beta_1 Academia_i + \beta_2 Cooperation_i + Controls + Y_i + \varepsilon \qquad (2)$$

Where $Novel_i$ represents the novelty score of each literature $i$. The independent variables $Academia_i$ and $Cooperation_i$ indicate whether the literature is from an academic or cooperative institution, respectively. The variable Controls includes a set of control variables, $Y_i$ denotes the publication year, and $\varepsilon$ represents the error term in the model.

To further refine the analysis, this study constructed a logistic regression model to verify the robustness of research conclusions and analyze whether institutional type correlates with literature being highly novel.

$$\ln\left(\frac{P(NS\ Top_i)}{1 - P(NS\ Top_i)}\right) = \alpha + \beta_1 Academia_i + \beta_2 Cooperation_i + Controls + Y_i \quad (3)$$

$P(NS\ Top_i)$ equals 1 indicates that the ith document belongs to the top novelty literature of that year. It should be noted that the formula does not include an explicit residual term, as logistic regression uses maximum likelihood estimation (MLE) for model fitting, rather than the ordinary least squares (OLS) method used in linear regression. Therefore, the error term is not directly included in the model expression.

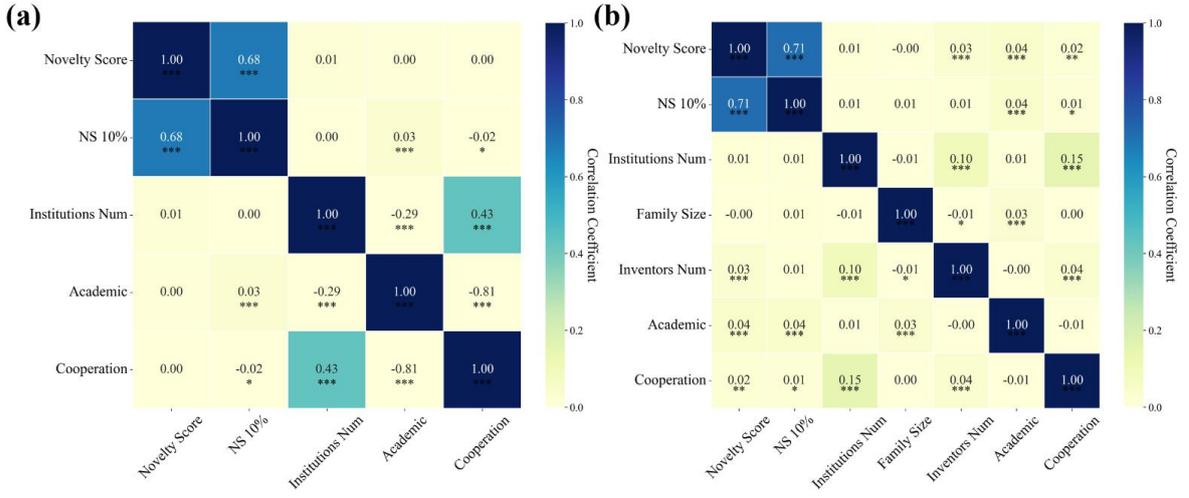

**Figure 2. Pearson correlation coefficient matrix (a) Correlation between variables in papers (b) Correlation between variables in patents**

However, OLS regression focuses on the effect of independent variables on the mean of the dependent variable, the logistic regression model concentrates on a single extreme outcome. To fully examine how institutional types shape novelty across the entire distribution of novelty scores, we further employed quantile regression. By estimating the conditional effects of independent variables at specific quantiles of the novelty score, this method complements the two aforementioned models. The model specification is as follows:

$$Q_\tau(Novel_i) = \alpha_\tau + \beta_{1\tau} Academia_i + \beta_{2\tau} Cooperation_i + Controls + Y_i \quad (4)$$

Where $Q$ denotes the quantile function, $\tau$ represents the target quantile (with $\tau \in (0,1)$), and $\beta_\tau$ indicates the conditional effect on the novelty at the $\tau - th$ quantile. Specifically, we examined the 25th, 50th, and 75th quantiles, which correspond to low, medium, and high novelty levels, to observe how different types of institutions perform across these heterogeneous novelty quantiles.

It can be observed that the distribution scales of the novelty scores for patents and papers exhibit significant differences in Table 2. Directly conducting a pooled analysis may lead to result biases due to scale inconsistency. To address this issue, this study first applies the z-score standardization method to separately transform the patent and paper data. By mapping the raw data to a standard normal distribution with a mean of 0 and a standard deviation of 1, the scale differences between different data sources are eliminated, enabling fair comparison and pooled analysis of the novelty scores of patents and papers under a unified standard scale. The transformation formula is as follows:

$$Z_{\text{Novelty},i} = \frac{Novelty_{i,\text{raw}} - \mu_{\text{Novelty}}}{\sigma_{\text{Novelty}}} \sim \mathcal{N}(0,1) \quad (5)$$

where $Novelty_{i,\text{raw}}$ represents the raw novelty score, $\mu_{\text{Novelty}}$ the population mean of this indicator, and $\sigma_{\text{Novelty}}$ the population standard deviation.

**Results**

This study analyses papers published between 2000 and 2022 in three representative conferences and patents filed with the USPTO, focusing on the novelty differences across three types of publishing institutions:

academia, industry, and collaboration. Our research compares the performance of different institution types in terms of novelty in literature and investigates the relationship between team size and novelty. The aim is to reveal how team size influences innovation across different types of scientific and technical literature.

*Trends in publication volume of papers and patents*

The field of NLP has experienced rapid growth, with a steady annual increase in patents and papers since 2000. The slight decrease in patent numbers in 2022 compared to 2021 is due to the America Invents Act (AIA), Section 35 U.S.C. § 122(b), which requires patents to be published 18 months after the earliest filing date, unless the applicant requests early publication. As of the retrieval date, some 2022 patents had not yet been published, which is common.

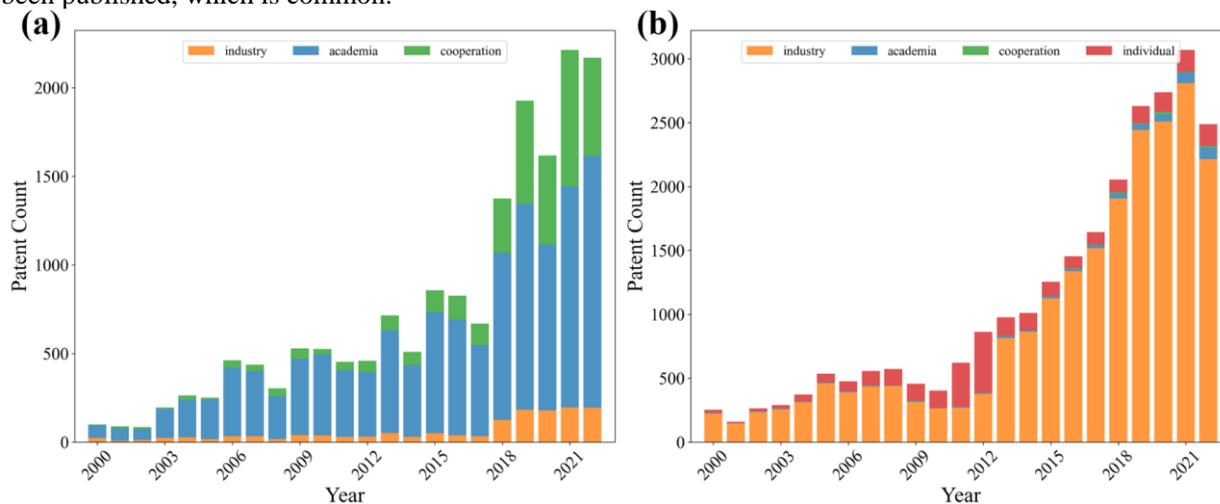

**Figure 3. Annual publication volume of papers and patents. (a) Annual publication volume of papers (b) Annual publication volume of patents**

In addition, the distribution of patent numbers across institutions is more uneven compared to papers, with specific proportions detailed in the previous section on institutional distribution. Despite the concentration of the world's top higher education resources in the United States and the majority of government research funding directed towards universities, university-originated patents account for less than 4% of the total national patents, with corporate patents dominating the majority, followed by individual applications[7]. This phenomenon also exists in the field of NLP, with the academic sector participating in significantly fewer patent applications than the industrial sector. The annual publication volume of papers and patents is shown in Figure 3.

*Novelty measurement results under a unified framework*

In this section, we address RQ1. We first use the entity recognition models discussed in previous chapters to extract fine-grained knowledge entities from each paper and patent. Then, we leverage the pre-trained SciBERT model to obtain semantic vectors for the entities in both patents and papers. We evaluate their novelty based on the semantic distance between different entities, with the overall distribution of entity semantic distance novelty shown in Figure 4.

---

[7] https://ncses.nsf.gov/pubs/nsb20204/invention-u-s-and-comparative-global-trends

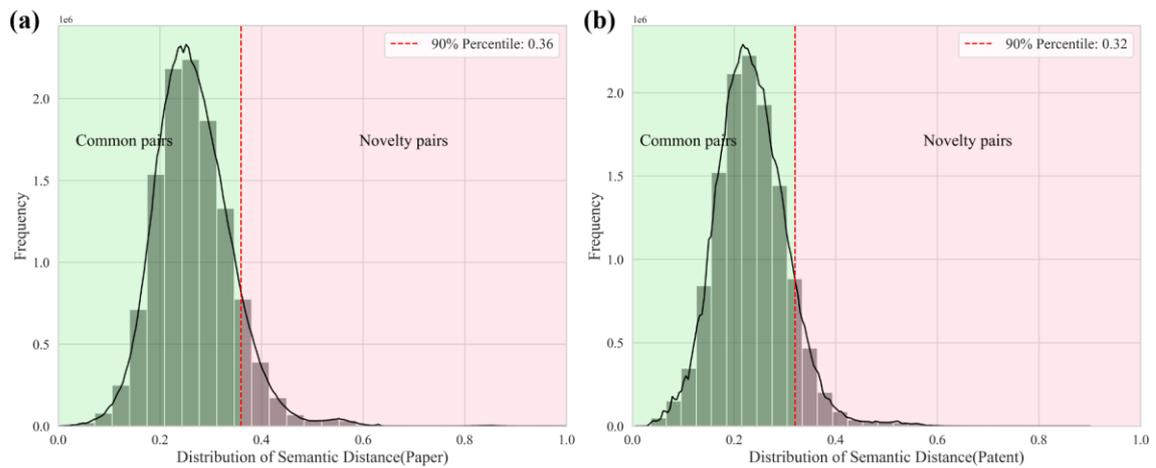

**Figure 4. Semantic distance distribution of fine-grained knowledge entities (a) Semantic distance distribution of paper entities; (b) Semantic distance distribution of patent entities.**

These semantic distances will be used to determine similarity. Specifically, we classify combinations into two types based on the semantic distance. Combinations are classified into Novelty (top 10%) and Common pairs by semantic distance. Final novelty depends on the proportion of Novelty pairs to total pairs.

To further explore the differences in the contribution of entity combinations to novelty between patents and papers, we analyze the average semantic distance of four entity combinations in the top 10% of papers and patents with the highest novelty. The results for papers are presented in Appendix Figure 8, and those for patents in Appendix Figure 9.

*Trends in novelty manifestations across different entity types*

In this section, we answer RQ2. Using the degree of atypicality in entity combinations as the core metric for novelty, we further analyzed the trends in semantic distance changes for entity combinations in papers and patents, as shown in Figure 5 and Figure 6. Given the study's focus on novelty, this section analyzes only the top 6 rankings from Appendix Figure 8 and Appendix Figure 9. Overall, as the field continues to evolve, the semantic distance between entities shows a general downward trend. This phenomenon reflects a narrowing gap in how patents and papers leverage prior knowledge, indicating increasingly seamless integration of knowledge within the domain (Park et al., 2023).

Specifically, regarding the Method-Metric combination identified as the primary innovation model in the field, Figure 5 reveals that there is virtually no difference between industry and academia in terms of semantic distance for Method-Metric entity combinations. Furthermore, no significant difference exists between the two sectors in the semantic distance of Method-Method combinations. These findings indicate that both industry and academia focus on innovation at the methodological level, and their innovation pathways consistently maintain a high degree of knowledge coherence within this dimension.

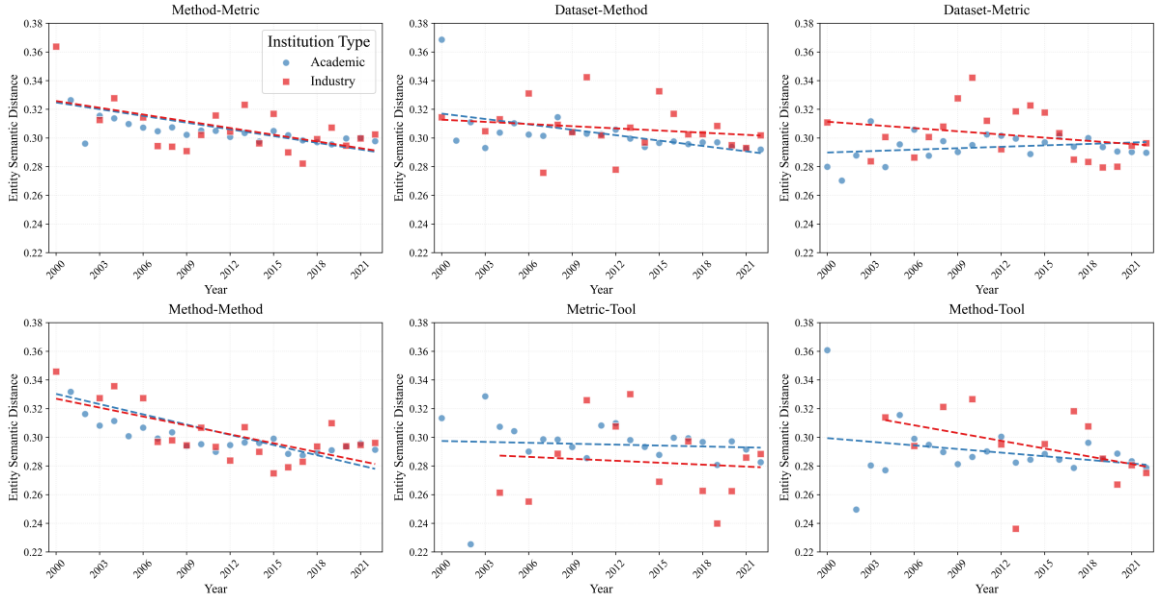

**Figure 5. Trends in semantic distance variations among different entity combinations in the paper**

In Dataset-related entity combinations, the industrial sector exhibits significantly higher semantic distance than the academic sector. This disparity highlights the industry's distinctiveness in dataset innovation, a characteristic closely linked to its access to larger-scale data resources (Ahmed et al., 2023). With such data scale advantages, the industrial sector can support forward-looking research. In contrast, the academic sector shows higher semantic distance than the industry in Tool-Metric combinations. The industry's lower semantic distance in this category may derive from its stronger focus on practical applications of mature technologies: this focus demands higher stability for Tool-Metric compatibility, fostering relatively fixed association patterns. Conversely, academic research, unconstrained by practical implementation scenarios, tends to explore innovative combinations of tools and metrics.

We subsequently analyzed the semantic distance trends of entity combinations within patents, with results presented in Figure 6. Due to the relatively limited number of patent disclosures in the academic domain, certain entity combinations did not emerge in the early stages.

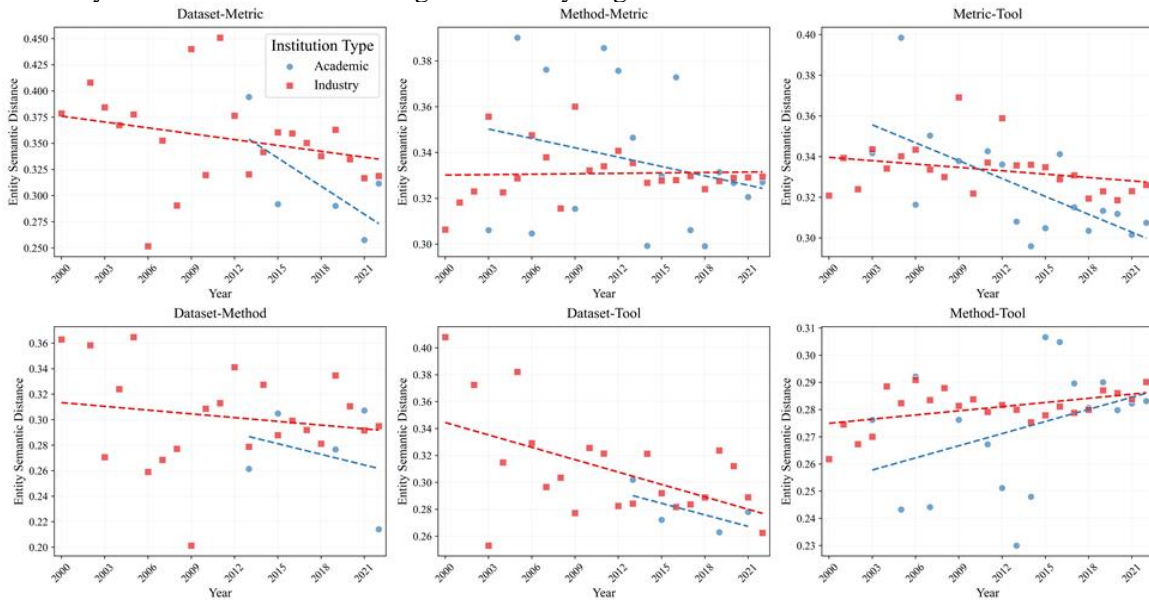

**Figure 6. Trends in semantic distance variations among different entity combinations in the patent**

For Dataset-related entity combinations (Dataset-Metric, Dataset-Method, and Dataset-Tool), the analysis revealed patterns similar to those observed in the paper research: across all three Dataset-related combinations, the industrial sector exhibited higher semantic distance compared to the academic sector. The consistent trends across both papers and patents clearly demonstrate that the industrial sector benefits not only from the existence of papers on datasets but also from their application in patent technology development. This indicates that the industry explores more boundary-pushing, dataset-driven innovation pathways in patent development. In terms of method entities, the two sectors exhibit distinct differences. For the Method-Metric combination, the academic domain demonstrates a higher semantic distance than the industrial domain. This finding indicates that the academic domain holds a relative advantage in the validation of methodological innovations. Conversely, for the Method-Tool combination, the advantage shifts to the industrial sector. The industry exhibits a higher semantic distance in the Method-Tool combination, reflecting its stronger capability to integrate methods with practical tools.

*Regression result of novelty differences across various types of institutions*

In this section, we answer RQ3. We conducted a series of multivariate regressions to compare the novelty of patents and papers across different institutional types. First, we show the summary analysis results of patents and papers, and then analyze patents and papers separately to observe their heterogeneity. In addition, quantile regression is further introduced to explore the heterogeneity effect under different novelty.

Table 3 presents the regression analysis results based on the full sample of scientific and technological literature. To eliminate differences in the original novelty value distributions between patents and papers, we first standardized the novelty values using Formula 4 to perform z-score normalization. This process aligned the novelty values of patents and papers onto the same distribution dimension.

**Table 3. Regression results for all scientific literature novelty**

| VARIABLES | Novelty Score (Z-score transformed) | | NS Top (TOP 10% OR NOT) | |
|---|---|---|---|---|
| | Model (1) | Model (2) | Model (3) | Model (4) |
| *Academic* | 0.127*** | 0.080*** | 0.518*** | 0.278*** |
| | (0.026) | (0.025) | (0.091) | (0.087) |
| *Cooperation* | 0.109*** | 0.079** | 0.347*** | 0.210* |
| | (0.029) | (0.032) | (0.105) | (0.110) |
| Ln(*Authors num*) | | 0.055*** | | 0.081*** |
| | | (0.010) | | (0.031) |
| Ln(*Institutions num*) | | 0.025 | | 0.150*** |
| | | (0.018) | | (0.056) |
| Ln(*Entities num*) | | -0.367*** | | -1.541*** |
| | | (0.013) | | (0.037) |
| Constant | -0.067 | 1.061*** | -2.389*** | 2.294*** |
| | (0.081) | (0.089) | (0.222) | (0.248) |
| Year fixed effects | Yes | Yes | Yes | Yes |
| Type fixed effects | Yes | Yes | Yes | Yes |
| Observations | 37,229 | 37,229 | 37,229 | 37,229 |
| R-squared | 0.005 | 0.033 | 0.002 | 0.068 |

**Note:** Robust standard errors in parentheses，*** p<0.01, ** p<0.05, * p<0.1

In the specific regression models, Models (1) and (2) employ z-score standardized novelty values as the core dependent variable, focusing on examining the impact of different participants on novelty. The regression results indicate that the coefficients for the core explanatory variables "Academic" and "Cooperation" both passed statistical significance tests, with positive signs for both coefficients. This finding indicates that within an analytical framework encompassing all scientific and technological literature, direct participation by the academic sector or participation through collaboration positively promotes the novelty level of outcomes.

Furthermore, the logistic regression results indicate that both in Model (3) without control variables and Model (4) with control variables, the effects of "Academic" and "Cooperation" remain statistically significant. Specifically, in Model (4) with control variables, the coefficient for "Academic" is 0.278, with an OR of $e^{0.278} = 1.32$; The coefficient for "Cooperation" is 0.210, OR is $e^{0.210} = 1.23$. This indicates that, controlling for other variables, academic institution participation increases the odds ratio of achieving high novelty by 32%, while academia collaboration participation increases this odds ratio by 23%.

In terms of model specification, this study not only controls for fixed effects by year to eliminate potential interference from temporal trends on novelty but also fully accounts for the inherent differences between patents and papers within the sample. Although the aforementioned z-score standardization and unified conceptual space construction have mitigated the heterogeneity between the two document types to some extent, to avoid potential confounding factors. This study additionally controlled for a fixed effect of document type in all regression models.

*Novelty heterogeneity between papers and patents*

Then, we analyze the heterogeneity of patents and papers. We first made a preliminary analysis of the novelty differences between papers and patents of different types of institutions, and the results are shown in Figure 7. Figure 7 (a) illustrates the distribution of novelty in the papers. Numerical results indicate that the mean novelty score for academic papers is 0.1013, while the industry sector averages 0.0987 and academia-industry collaboration papers average 0.1015. These three groups exhibit closely aligned novelty levels. Notably, the distribution patterns reveal that academic papers exhibit greater dispersion, featuring a small number of extremely novel values. This suggests academic institutions may possess greater potential for innovation in their papers, with a higher probability of producing highly novel research.

Figure 7 (b) presents the results for patents. The mean novelty score for academic patents is 0.1373, while the industry average is 0.1146, indicating that academic institutions significantly outperform industry in novelty. The mean number of collaboration types is 0.1427. Compared to research papers, academic institutions produce more highly novel outcomes in patent innovation, with a mean novelty score markedly higher than that of industry.

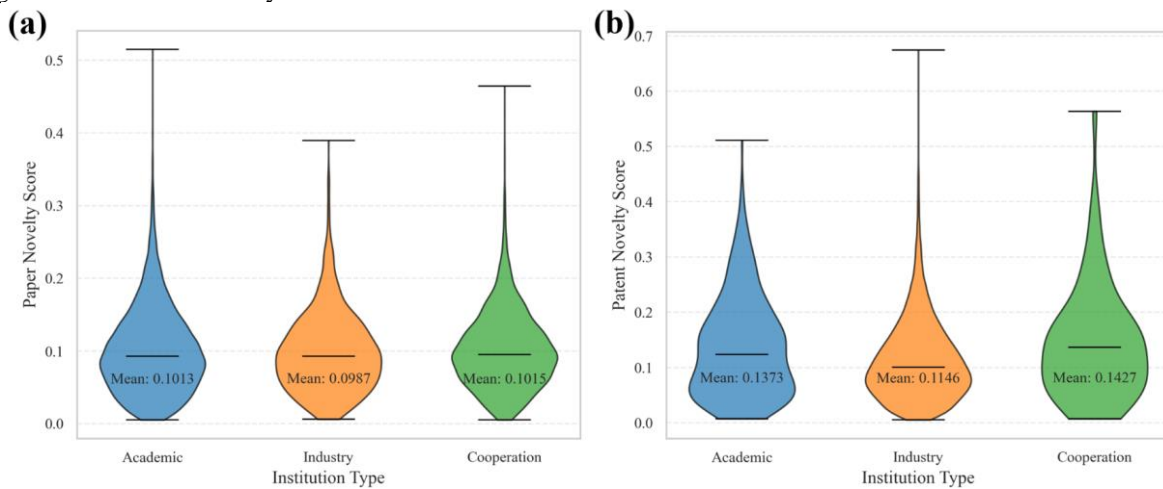

**Figure 7. Violin Plots of novelty distribution. (a) Novelty differences across publishing institutions in the papers (b) Novelty differences across publishing institutions in the patents**

Further regression results of the paper are shown in Appendix Table 6. In the ordinary linear regression Model (5) that only includes institution type, the coefficient for "Academia" is 0.003, which is statistically significant at the 5% level ($\beta = 0.003$, $p < 0.05$). This indicates that the average novelty score of academic papers is 0.003 higher than that of industrial papers. The coefficient for "Cooperation" is also 0.003, but the difference is not significant. This suggests that although the novelty of collaboration papers is numerically higher than that of industrial papers, this difference lacks statistical significance. After incorporating control variables into Model (6), the coefficient for "Academia" decreases to 0.002 and becomes insignificant ($p = 0.168 > 0.1$). The novelty gap between academic and industrial papers is no longer statistically significant.

We supplemented the quantile regression analysis (see Appendix Table 8 for the results), found that the coefficient of academic variables increased with the increase of the quantile and was statistically significant, indicating that there were differences in the performance of different novelty intervals: the coefficient was ($\beta = 0.002, 0.001$) at the 25% quantile (low novelty) was no significant difference with the industry. The coefficient at 75% quantile (high novelty) was ($\beta = 0.007, 0.004$) and significant ($p < 0.1$, $p < 0.01$). This shows that the performance of academia is similar to that of industry in conventional research, but it has obvious advantages in highly novel research.

Therefore, we further analyzed the differences in the possibility of becoming a highly novel paper. Results from the logistic regression Model (3) show that the coefficient for "Academia" is 0.331 ($\beta = 0.331$, $p < 0.01$). Converting this to an odds ratio ($OR = e^{0.331} \approx 1.39$) indicates that, without controlling for other variables, academic papers are 1.39 times more likely to be high novelty outputs than industrial papers. The coefficient for "Cooperation" in Model (7) is 0.158, which remains insignificant. This aligns with findings from studies such as Liang et al. (2024), who noted that "academic–industry collaborations struggle to replicate the novelty of academic teams and tend to look similar to industry teams".

After adding all control variables to Model (8), the coefficient for "Academia" decreases to 0.204, with an odds ratio of $(OR = e^{0.204} \approx 1.23)$. This indicates that even when accounting for confounding factors, the odds of academic papers being highly novel are still 23% higher than those of industrial papers. It retains marginal statistical significance ($p < 0.1$), validating academia's relative advantage. The coefficient for "Cooperation" in Model (4) drops to 0.057 ($OR = e^{0.057} \approx 1.06$), meaning collaboration papers have a limited likelihood of being high novelty, with no substantial statistical difference from industrial papers.

In summary, there is no significant difference in the average novelty score between academia and industry. However, academia has a more prominent advantage in high novelty outputs. As for academia-industry collaborative papers, their novelty is only slightly higher than that of industry papers in numerical terms, but such a difference lacks statistical significance across all models.

Appendix Table 7 presents the regression results for patents. Model (9) includes only institutional type and year fixed effects. The coefficient for "Academia" is 0.022 ($p<0.01$), indicating that the average novelty score of patents from academic institutions is significantly higher than that of patents from industrial institutions by 0.022. The coefficient for "Cooperation" is 0.026 ($p<0.05$), with an average novelty score also significantly higher than industrial institution patents by 0.026 units. After introducing control variables in Model (10), the coefficient for "Academia" remained statistically significant at the 1% level, while the "Cooperation" coefficient decreased to 0.022 ($p<0.1$). This indicates that although the novelty advantage of patents from collaborative institutions did not completely disappear, it weakened significantly after controlling for other variables.

Further analysis of the quantile regression results (see Table 9 in the appendix) reveals that as the quantile rises from 25% to 75%, the coefficient increases consistently. This indicates that, similar to the performance observed in papers, academia also demonstrates a more prominent advantage in high novelty within the patent domain. In addition, the results of the logistic regression Model (11) results indicate that the coefficient for "Academic" is 0.762 and significant at the 1% level ($\beta = 0.762$, $p<0.01$). Converting this to an odds ratio ($OR = e^{0.762} \approx 2.14$) reveals that, without controlling for other variables, patents from academic institutions have an advantage in high novelty over patents from industrial institutions by a factor of 2.14. The coefficient for "Cooperation" was 0.695, significant at the 5% level ($\beta = 0.695$, $p<0.05$).

After incorporating all control variables in Model (12), the coefficient for "Academic" decreased to 0.369 but remained significant at the 1% level ($\beta =0.369$, $p<0.01$), and $OR$ is $e^{0.369} \approx 1.45$. At this point, the coefficient for "Cooperation" decreased to 0.502 ($\beta=0.502$, $p>0.1$). This indicates that while there is a numerical difference, this difference is not statistically significant.

We observed the effect of control variables, and the coefficient for the number of IPC classification codes was ($\beta = 0.005, 0.253$, $p<0.01$), indicating that higher diversity in the knowledge domains covered by patents is more likely to drive cross-domain innovation, thereby enhancing patent novelty. The coefficient for patent family size was -0.004 and -0.100 ($p<0.01$), suggesting that larger patent families tend to focus on incremental improvements rather than breakthrough innovations, thus reducing novelty. The performance of entity number aligns with Fleming and Sorenson's (2001) theory, where the marginal benefit of increasing component quantity diminishes for patent enhancement. Given that novelty in this study is calculated based on the proportion of novel entities, it maintains a negative value in the regression model. Furthermore, the addition of control variables significantly improved the R-squared value for all models, confirming that incorporating these variables enhanced the models' explanatory power.

**Discussion**

This study adopted fine-grained knowledge entity analysis to evaluate the novelty of patents and papers within the NLP field. Through regression analysis, the level of novelty in academia surpasses that in industry when considering both papers and patents. This finding is consistent with the results of previous work (Chen et al., 2024; Liang et al., 2024). Regarding innovation focus, both academia and industry emphasize methodological innovation in papers. Industry holds a unique advantage over academia in terms of dataset resources. Notably, the academic sector demonstrates a more pronounced advantage in producing highly novel outcomes. Regarding collaborative research outputs, academic–industry collaborations struggle to replicate the novelty of academic teams and tend to resemble the work of industry teams (Liang et al., 2024). As a catalyst, academia significantly promotes the enhancement of novelty, both in terms of filing patents individually and participating in collaborative patent research.

*Implications*

The theoretical implications of this study lie in three aspects. First, by transferring knowledge from entity recognition models in the academic paper domain to the patent entity recognition task and integrating an entity-based novelty measurement approach, we achieved a unified quantitative assessment of novelty across both patent and paper. This establishes a viable framework for subsequent cross-level novelty evaluation between papers and patents on larger datasets.

Second, to address the divergence in novelty perceptions between academia and industry in existing research, this study integrates top conference papers with patent data to provide new empirical evidence. The findings reveal novelty differences between patents and papers, with knowledge entities across different scientific literatures also exhibiting innovation disparities. At the paper level, the overall novelty gap between academia and industry is relatively small, particularly in Method-based innovation domains, where the difference is negligible. This is attributable to the rigorous peer-review mechanisms of the three major conferences in natural language processing. Concurrently, the industry leverages its inherent dataset advantages to conduct extensive innovation activities.

Third, this study enriches the empirical foundation of Recombinant Search Theory (Fleming, 2001; Fleming & Sorenson, 2001) at the entity level: It refines the core perspective that technological innovation originates from component local search and recombination (IPC subclass) to the micro-level of knowledge entities. Building upon Aceves and Evans' (2024) conceptual space theory, this study aligns paper and patent measurement dimensions through a unified word embedding model, thereby enriching the analytical dimensions of component combinations in technological innovation research. Within this framework, this study similarly identifies a diminishing marginal returns trend in output associated with increasing component numbers. This phenomenon stems partly from cognitive constraints limiting the full utilization of multiple components (Fleming & Sorenson, 2001). It may also result from reduced innovation due to the narrower application of prior knowledge (Park et al., 2023).

This study also has clear practical value. Our research reveals that academia demonstrates higher overall novelty outputs, which aligns with the view of Brescia et al. (2016), that the openness and collaborative environment of academia are more conducive to fostering new ideas and supporting interdisciplinary innovation. The findings of this study facilitate the formulation of differentiated incentives for academia-industry collaboration for policymakers. At the patent innovation level, greater support should be provided for joint patent applications between industry and academia. This could be achieved through measures such as subsidizing patent application costs and establishing priority review channels, thereby fully leveraging academia's role in enhancing patent novelty. At the paper innovation level, avoid mandating collaboration models. Instead, prioritize supporting independent university teams in conducting fundamental research while guiding enterprises to participate in paper research through data sharing rather than joint authorship, thereby enhancing innovation effectiveness.

*Limitations and future works*

Although this study has empirically revealed the novelty differences between academia and industry through a unified semantic space, it nevertheless has several limitations. First, the research sample is constrained in scope. The analyzed papers are limited to three representative NLP conferences, excluding NLP research published in other relevant conferences and journals. Patents are restricted to the core CPC class G06F40, omitting related patent activities in adjacent CPC classes critical to NLP (e.g., machine learning G06N20, information retrieval G06F16). While this sampling strategy guarantees highly relevant data, it may underestimate the actual breadth of NLP-related research papers and patent activities. Second, new words in the NLP field iterate quickly, which may trigger dynamic fluctuations in semantics. While entity normalization versus annual fixation effects can mitigate this situation, it can still introduce fine bias to the novelty measure of the entity combination. Third, the study focuses solely on NLP. As a rapidly emerging field with deep engagement across academia, industry, and research, other domains may exhibit heterogeneity due to differences in development pace, institutional participation models, or knowledge production logic. Thus, caution should be exercised when generalizing the research conclusions to other domains. Finally, although this study employs an analytical approach grounded in classical combinatorial innovation theory (Liu et al., 2022), it currently lacks a gold-standard validation process, which constitutes a challenge pervasive in contemporary relevant research.

Future research may advance in three directions: First, expand the scope of research data. For papers, future studies can extend the data coverage to a wider range of NLP-related conferences and journals. For patents, more refined classification and filtering approaches can be employed to identify and incorporate patents from adjacent technical fields, thereby comprehensively capturing patent activities in NLP. Second, it can overcome the limitations of relying solely on semantic distance metrics to measure novelty, exploring a more comprehensive and multidimensional innovation evaluation framework. Finally, supplement a validation process based on gold standards. By integrating expert review opinions, systematically test the model's effectiveness in distinguishing documents with specific novelty levels, further enhancing the rigor and reliability of the research outcomes.

**Acknowledgment**

This paper was supported by the National Natural Science Foundation of China (Grant No.72074113) and Jiangsu Province Graduate Student Research Practice Innovation Program (Grant No. KYCX25_0820). This paper is an extended version of the ISSI 2025 paper "Zhao, H., Zhao, Y., & Zhang, C. (2025). Exploring novelty differences between industry and academia: A knowledge entity-centric perspective. *20th International Conference on Scientometrics & Informetrics*. Yerevan, Armenia. https://doi.org/10.51408/issi2025_088".

**Declarations**

Conflict of interest: The author(s) declared no potential conflicts of interest with respect to the research, authorship, and/or publication of this article.

**Appendix**

**Table 4. Top 5 entities in four types extracted from papers and patents**

| Type | Paper | | Patent | |
|---|---|---|---|---|
| | *Entity* | *Frequency* | *Entity* | *Frequency* |
| Method | BERT | 4160 | Neural network | 3021 |
| | LSTM | 3565 | Machine learning | 1608 |
| | Attention Mechanism | 3392 | N-gram | 1365 |
| | Transformer | 3321 | Language models | 1160 |
| | N-gram | 3287 | Deep learning | 960 |
| Tool | Pytorch | 742 | Computer system | 11646 |
| | MOSES | 654 | Storage medium | 10413 |
| | GIZA++ | 54 | User interface | 9323 |
| | Python | 438 | Computer program | 8738 |
| | NLTK | 397 | Operating system | 7636 |
| Dataset | Wikipedia | 3542 | Emoji | 306 |
| | WordNet | 2152 | Email | 122 |
| | Twitter | 1210 | Social media | 86 |
| | Wall Street Journal | 1006 | World wide web | 67 |
| | Amazon Mechanical Turk | 982 | Twitter | 43 |
| Metric | Accuracy | 10696 | Accuracy | 5278 |
| | $F_1$ | 6956 | Confidence | 2500 |
| | Precision | 6075 | Efficiency | 2195 |
| | Recall | 5578 | Relevance | 1612 |

| BLEU | 3264 | Error | 1453 |

**Table 5. Prompt for classifying institution names in the list**

I am providing a list of names that may include individuals, academic institutions (e.g., universities, research institutes), or industrial organizations (e.g., companies, corporations). Please review each name and classify it into one of the following categories:
"Academia" if it refers to a university, college, research institute, or other research organization with an education focus.
"Industry" if it refers to a company, corporation, enterprise, or business entity.
"Individual" if it appears to be a personal name (not an institution).
Do not provide rationales for classifications. If a name is ambiguous or cannot be clearly categorized, label it as "Other".

*Results of semantic similarity for different entity combination types*

As shown in Appendix Figure 8, combinations of entities within the same category, such as tools, metrics, and datasets, have relatively high semantic similarity. This indicates that these combinations are closer in semantic distance and contribute relatively little to novelty. The top-ranked combinations are Method-Metric, Dataset-Method, Dataset-Metric, and Method-Method. Among the core combinations mentioned above, Method-Metric exhibits the lowest semantic similarity (highest semantic distance). In fact, method design and metric definition share a deeply intertwined relationship: the introduction of new methods must be accompanied by corresponding metrics for validity assessment, and the "method validation logic" represents a typical pathway for producing highly novel research outcomes. Additionally, the Method-Dataset pairing ranks prominently. High-quality datasets form a crucial foundation for methodological innovation. Research by Pramanick et al. (2025) confirms this: first, new datasets tend to attract higher citation rates; second, papers introducing new methods also garner increased citation attention.

Appendix Figure 9 presents patent entity combination similarity. Among these, same-type entity pairs (Dataset-Dataset, Tool-Tool, Method-Method, Metric-Metric) exhibit the highest semantic similarity, which aligns with Fleming and Sorenson's (2001) technology landscape recombination search theory. When inventors adopt a modular strategy for patent development, while it reduces uncertainty in technological restructuring and enhances invention success rates, it simultaneously narrows the space for cross-module collaborative innovation, ultimately limiting breakthrough inventions. For high novelty patents, their core entity combinations are Dataset-Metric, Method-Metric, Metric-Tool, and Dataset-Method, sharing similarities with the entity combination patterns of papers discussed earlier. However, patents and papers differ significantly in research focus: patents rarely use data and metrics to verify technical validity; instead, they demonstrate value by describing technical features and achieved effects. Nevertheless, combining datasets with evaluation metrics or integrating datasets with methods still generates novel outcomes. Furthermore, patents emphasize practical applications with higher engagement from Tool-based entities, where certain Tool-Metric combinations emerge as key pathways for technological innovation.

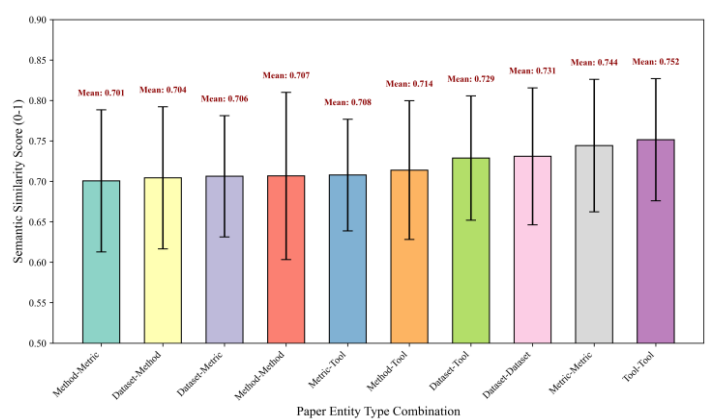

**Figure 8. Semantic similarity differences between paper entities**

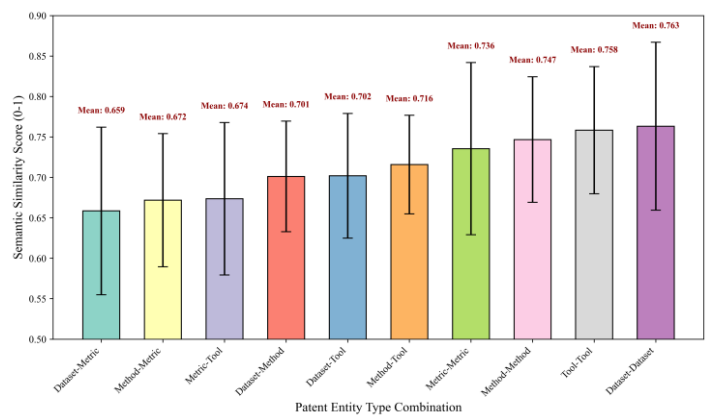

**Figure 9 Semantic similarity differences between patent entities**

*Results of linear and logistic regression*

**Table 6. Regression results for paper novelty**

| VARIABLES | Novelty Score (Continuous variable) | | NS Top (TOP 10% OR NOT) | |
|---|---|---|---|---|
| | Model (5) | Model (6) | Model (7) | Model (8) |
| *Academic* | 0.003** | 0.002 | 0.331*** | 0.204* |
| | (0.002) | (0.002) | (0.105) | (0.108) |
| *Cooperation* | 0.003 | 0.001 | 0.158 | 0.057 |
| | (0.002) | (0.002) | (0.115) | (0.128) |
| Ln(*Authors num*) | | -0.001 | | -0.098 |
| | | (0.001) | | (0.061) |
| Ln(*Institutions num*) | | 0.003** | | 0.194*** |
| | | (0.001) | | (0.065) |
| Ln(*Entities num*) | | -0.009*** | | -1.153*** |
| | | (0.001) | | (0.067) |
| Constant | 0.121*** | 0.144*** | -2.185*** | 0.750 |
| | (0.015) | (0.016) | (0.491) | (0.517) |
| Year fixed effects | Yes | Yes | Yes | Yes |
| Observations | 16,295 | 16,295 | 16,295 | 16,295 |
| R-squared | 0.005 | 0.009 | 0.002 | 0.029 |

**Note:** Robust standard errors in parentheses, *** $p<0.01$, ** $p<0.05$, * $p<0.1$

**Table 7. Regression results for patent novelty**

| VARIABLES | Novelty Score (Continuous variable) | | NS Top (TOP 10% OR NOT) | |
|---|---|---|---|---|
| | Model (9) | Model (10) | Model (11) | Model (12) |
| *Academic* | 0.022*** | 0.013*** | 0.762*** | 0.369*** |
| | (0.004) | (0.004) | (0.125) | (0.138) |
| *Cooperation* | 0.026** | 0.022* | 0.695** | 0.502 |
| | (0.013) | (0.012) | (0.333) | (0.352) |
| Ln(*Authors num*) | | 0.005*** | | 0.121*** |
| | | (0.001) | | (0.038) |
| Ln(*Institutions num*) | | -0.002 | | -0.075 |
| | | (0.003) | | (0.142) |
| Ln(*Entities num*) | | -0.036*** | | -1.906*** |
| | | (0.001) | | (0.049) |
| Ln(*IPC num*) | | 0.005*** | | 0.253*** |
| | | (0.001) | | (0.053) |
| Ln(*Family size*) | | -0.004*** | | -0.100** |

|  | | | (0.001) | | | (0.043) |
|---|---|---|---|---|---|---|
| Constant | | 0.113*** | 0.217*** | -1.986*** | | 3.342*** |
| | | (0.005) | (0.006) | (0.224) | | (0.265) |
| Year fixed effects | | Yes | Yes | Yes | | Yes |
| Observations | | 20,934 | 20,934 | 20,934 | | 20,934 |
| R-squared | | 0.007 | 0.064 | 0.003 | | 0.111 |

**Note:** Robust standard errors in parentheses. *** p<0.01, ** p<0.05, * p<0.1.

*Results of quantile regression*

**Table 8. Regression results of novelty scores of papers under different quantiles**

|  | Novelty Score Q (0.25) | | Novelty Score Q (0.50) | | Novelty Score Q (0.75) | |
|---|---|---|---|---|---|---|
| VARIABLES | Model (13) | Model (14) | Model (15) | Model (16) | Model (17) | Model (18) |
| *Academic* | 0.001 | 0.002 | 0.002 | 0.002 | 0.007*** | 0.004* |
|  | (0.002) | (0.002) | (0.002) | (0.002) | (0.002) | (0.002) |
| *Cooperation* | 0.003 | 0.001 | 0.002 | 0.002 | 0.004* | 0.003 |
|  | (0.002) | (0.002) | (0.002) | (0.002) | (0.002) | (0.003) |
| Ln(*Authors num*) |  | 0.001 |  | 0.002 |  | -0.002 |
|  |  | (0.001) |  | (0.001) |  | (0.002) |
| Ln(*Institutions num*) |  | 0.001 |  | 0.001 |  | 0.003* |
|  |  | (0.001) |  | (0.001) |  | (0.002) |
| Ln(*Entities num*) |  | 0.016*** |  | -0.000 |  | -0.021*** |
|  |  | (0.001) |  | (0.002) |  | (0.002) |
| Constant | 0.037*** | -0.007 | 0.101*** | 0.086** | 0.205*** | 0.265*** |
|  | (0.010) | (0.006) | (0.035) | (0.036) | (0.021) | (0.020) |
| Year fixed effects | Yes | Yes | Yes | Yes | Yes | Yes |
| Observations | 16,295 | 16,295 | 16,295 | 16,295 | 16,295 | 16,295 |
| R-squared | 0.018 | 0.025 | 0.007 | 0.007 | 0.004 | 0.012 |

**Note:** Robust standard errors in parentheses. *** p<0.01, ** p<0.05, * p<0.1.

**Table 9. Regression results of novelty scores of patents under different quantiles**

|  | Novelty Score Q (0.25) | | Novelty Score Q (0.50) | | Novelty Score Q (0.75) | |
|---|---|---|---|---|---|---|
| VARIABLES | Model (19) | Model (20) | Model (21) | Model (22) | Model (23) | Model (24) |
| *Academic* | -0.001 | 0.001 | 0.021*** | 0.016** | 0.043*** | 0.019*** |
|  | (0.004) | (0.004) | (0.007) | (0.007) | (0.010) | (0.005) |
| *Cooperation* | 0.005 | -0.001 | 0.031*** | 0.019 | 0.024** | 0.021 |
|  | (0.011) | (0.013) | (0.010) | (0.022) | (0.011) | (0.017) |
| Ln(*Authors num*) |  | 0.004*** |  | 0.004*** |  | 0.006*** |
|  |  | (0.001) |  | (0.001) |  | (0.001) |
| Ln(*Institutions num*) |  | 0.001 |  | -0.003 |  | -0.005 |
|  |  | (0.003) |  | (0.003) |  | (0.005) |
| Ln(*Entities num*) |  | 0.003*** |  | -0.026*** |  | -0.059*** |
|  |  | (0.001) |  | (0.001) |  | (0.001) |
| Ln(*IPC num*) |  | 0.002** |  | 0.005*** |  | 0.007*** |
|  |  | (0.001) |  | (0.001) |  | (0.001) |
| Ln(*Family size*) |  | -0.003*** |  | -0.004*** |  | -0.006*** |
|  |  | (0.001) |  | (0.001) |  | (0.001) |
| Constant | 0.054*** | 0.045*** | 0.103*** | 0.178*** | 0.154*** | 0.336*** |
|  | (0.003) | (0.005) | (0.010) | (0.005) | (0.011) | (0.007) |
| Year fixed effects | Yes | Yes | Yes | Yes | Yes | Yes |
| Observations | 20,934 | 20,934 | 20,934 | 20,934 | 20,934 | 20,934 |
| R-squared | 0.003 | 0.006 | 0.005 | 0.023 | 0.007 | 0.073 |

**Note:** Robust standard errors in parentheses. *** p<0.01, ** p<0.05, * p<0.1.

*The relationship between novelty and citations*

In the view of Franceschini et al. (2012), citations in papers primarily reflect scholarly influence, whereas patent forward citations possess dual attributes as both vehicles for knowledge flow and indicators of technological application value.

Existing research has extensively examined citation patterns of highly novel outcomes: most studies indicate that increased patent novelty typically correlates with more forward citations (Jeon et al., 2022; Liu et al., 2025). However, it must be clarified that citation metrics measure ex post impacts after publication, not the intrinsic novelty of the work. Contrary perspectives exist. Arts & Fleming (2018) found that while exploring novel domains enhances invention novelty, insufficient knowledge reserves and high learning costs may diminish immediate value. Wang et al. (2017) found that highly novel research exhibits greater citation variance and often struggles to achieve high citation counts during its initial publication phase. Liang et al. (2023) further supplemented that highly novel outcomes typically face longer review and revision cycles, reflecting a "delayed recognition" bias in academic evaluation that resonates with citation lag phenomena.

To verify the robustness of novelty-citation associations, we employed the Mann-Whitney U test to compare citation outcomes between high novelty and common papers/patents. Drawing on five-year citation data for 15,871 papers from OpenAlex and forward citation data for 20,934 patents from proprietary databases, our analysis revealed:

a) For papers: High novelty samples exhibited significantly lower five-year citation counts, with this pattern being statistically robust across both the overall sample and academia subgroup ($p<0.001$; see Appendix **Figure 10**);

b) For patents: High novelty samples received significantly more forward citations, a consistent trend validated in both the overall patent sample and industry subgroup ($p<0.001$; see Appendix **Figure 11**).

These findings align with existing literature: Patent novelty is positively correlated with subsequent technological impact, as novel patents exhibit more active technology diffusion and greater average technical influence. In contrast, the association between paper novelty and citation performance remains negative or uncertain. This stems from high novelty papers being susceptible to systematic biases in evaluation, hindering their ability to gain widespread scholarly attention and recognition in the short term.

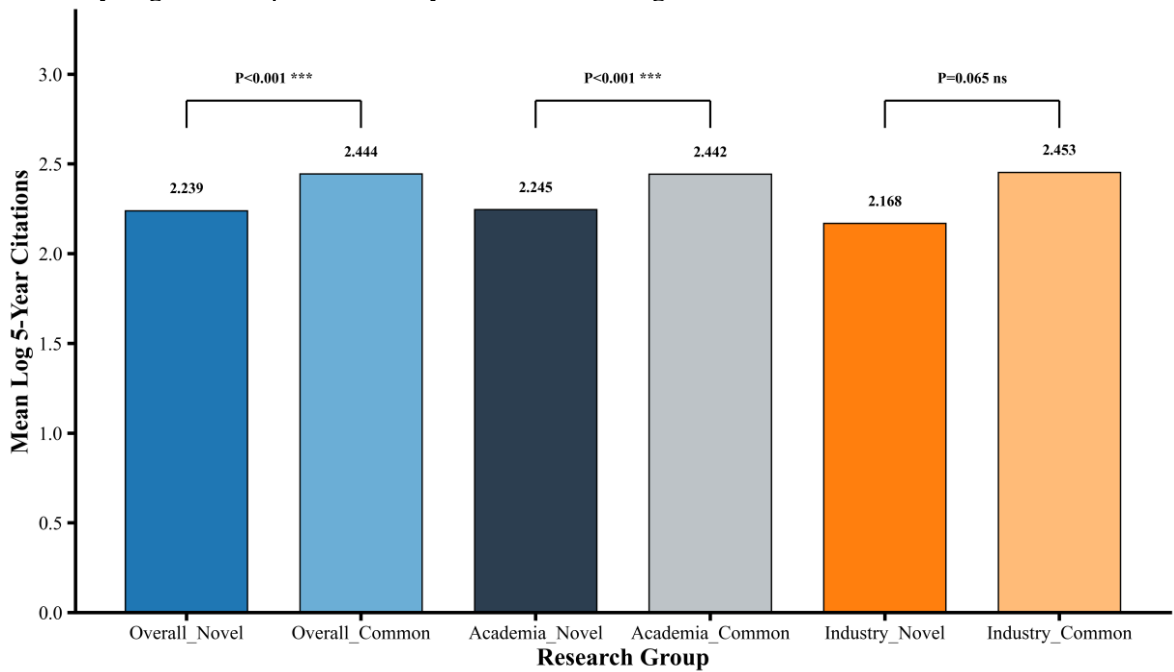

**Figure 10. Differences in 5-year citations between novel and common papers**

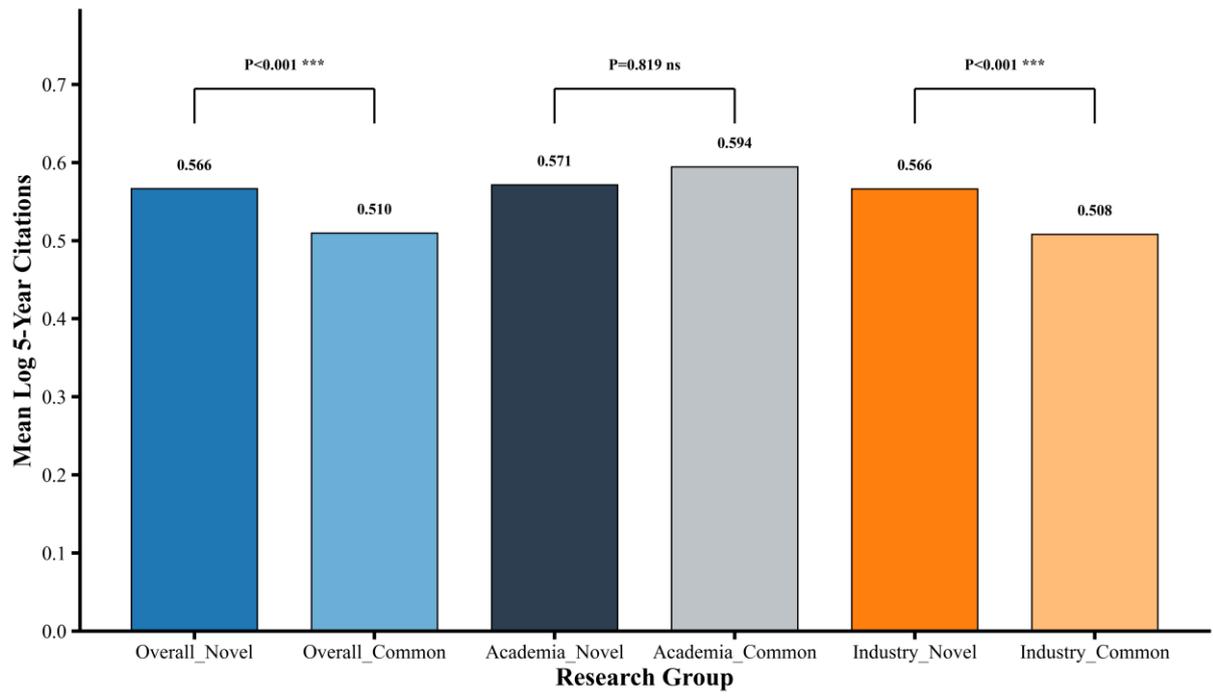

**Figure 11. Differences in 5-year citations between novel and common patents**